\documentclass{svmult}

\usepackage{makeidx}         
\usepackage{graphicx}        
\usepackage{multicol}        
\usepackage[bottom]{footmisc}

\usepackage{amssymb}
 
\makeindex             

\begin{document}

\title*{Tick size and price diffusion}
\author{Gabriele La Spada\inst{1}
\and
J. Doyne Farmer \inst{2}\and
Fabrizio Lillo\inst{3}}
\institute{Department of Economics, Princeton University, Princeton, NJ 08544-1021 USA,
Dipartimento di Scienze Economiche e Aziendali
LUISS
Viale Romania 12
I-00197 Roma, Italy
\texttt{gla@princeton.edu}
\and Santa Fe Institute, 1399 Hyde Park Road, Santa Fe, NM 87501, USA \texttt{jdf@santafe.edu}
\and Dipartimento di Fisica e Tecnologie Relative, Universit\`a di Palermo, 
Viale delle Scienze, I-90128, Palermo, Italy
Santa Fe Institute, 1399 Hyde Park Road, Santa Fe, NM 87501 USA \texttt{lillo@unipa.it}}
%
%
\maketitle

\section{Introduction}
\label{sec:1}

A tick size is the smallest increment of a security price. Tick size is typically regulated by the exchange where the security is traded and it may be modified either because the exchange enforces an overall tick size change or because the price of the security is too low or too high. There is an extensive literature, partially reviewed in Section 2 of the present paper, on the role of tick size in the price formation process. However, the role and the importance of tick size has not been yet fully understood, as testified, for example, by a recent document of the Committee of European Securities Regulators (CESR) \cite{Cesr10}. 

Tick size can affect security price in direct and indirect ways. It is clear that at the shortest time scale on which individual orders are placed the tick size has a major role which affects where limit orders can be placed, the bid ask spread, etc. This is the realm of market microstructure and in fact there is a vast  literature on the role of tick size on market microstructure. 
However, tick size can also affect price properties at longer time scales, and relatively less is known about the effect of tick size on the statistical properties of prices. The rationale is that since market microstructure affects price diffusion, if tick size affects microstructure it is likely  to also affect  price diffusion.  This point of view is strengthened  by the observation that there is growing evidence that microstructural events are important factors in explaining longer term price fluctuations \cite{bfl}. For example, large fluctuations in price returns are observed also at the individual transaction scale \cite{Farmer04} and their effects last for long periods \cite{Ponzi09}. 
Nonetheless the relation between market microstructure and price diffusion on a longer time scale is still not fully understood. It is therefore worth investigating the effect of tick size on price diffusion properties.

The present paper is divided in two parts. In the first (Section 2) we review the effect of tick size change on the market microstructure and the diffusion properties of prices. The second part (Section 3) presents original results obtained by investigating the tick size changes occurring at the New York Stock Exchange (NYSE). We show that tick size change has three effects on price diffusion. First, as already shown in the literature, tick size affects price return distribution at an aggregate time scale. Second, reducing the tick size typically leads to an increase of volatility clustering. We give a possible mechanistic explanation for this effect, but clearly more investigation is needed to understand the origin of this relation. Third, we explicitly show that the ability of the subordination hypothesis in explaining fat tails of returns and volatility clustering is strongly dependent on tick size. While for large tick sizes the subordination hypothesis has significant explanatory power, for small tick sizes we show that subordination is not the main driver of these two important stylized facts of financial market. Finally Section 4 concludes.

\section{Literature review}
\label{sec:2}
In this section we review some literature on the effect of tick size on market microstructure and on price diffusion. Most of the studies we consider are case studies of tick size changes in different markets. This section is divided in two parts. In the first we review the effect of tick size on market microstructure, while in the second we discuss  recent papers on the effect of tick size on price diffusion.

\subsection{Tick size and market microstructure}


Crack \cite{Crack94} and Ahn et al. \cite{Ahn96} studied the impact of the September 3, 1992 AMEX reduction in the minimum tick size from 1/8 to 1/16 for stocks priced under five dollars. They found  approximately a 10\% decline in quoted spreads and depths in addition to an increase in average daily trading volume of 45-55 \%.
Niemeyer and Sand\aa s \cite{Niemeyer94}  studied the Stockholm Stock Exchange and found that tick size is positively correlated to spread and market depth, and negatively correlated to volume.
Angel \cite{Angel97} found that small tick size narrows the bid-ask spread, but diminishes liquidity by making limit order traders and market makers more reticent to supply shares.

A series of studies \cite{Bacidore97,Porter97,Huson97,Ahn98} investigated the April 15, 1996 Toronto Stock Exchange (TSE) reduction in the minimum tick size to five cents. They found a significant decline in the quoted bid-ask spreads of 17-27\% and in the quoted depth of 27-52\%, while average trading volume displayed no statistically significant increase.

Bessembinder \cite{Bessembinder97} studied the Nasdaq stocks whose price level passes through \$10, and thus changed tick size from 1/8 to 1/32 and found that the effective spread fell by 11\%.
Ronen and Weaver \cite{Ronen98} studied the May 7, 1997 switch to 1/16 at  the AMEX and found reduced quote spreads and depths. They concluded that the implemented reduction to the minimum tick size has decreased transaction costs and increased liquidity.
Bollen and Whaley \cite{Bollen98} and Ricker \cite{Ricker98} studied the 1997 tick size reduction from 1/8 to 1/16  at the NYSE and found that volume weighted bid-ask spread declined by approximately  13-26\%
while quoted depth decreased between 38\% and 45\%. They concluded that the NYSE tick size reduction has improved the liquidity of the market especially for low-priced shares.
Jones and Lipson \cite{Jones01} used institutional data to study the effect of tick size changes at the Nasdaq and NYSE. They found that trading costs decreased for smaller trades, but increased for larger trades.

Goldstein and Kavajecz \cite{Goldstein00} studied the tick size change from 1/8 to 1/16 at the NYSE. They found that the quoted bid-ask spread narrowed by 14.3\% (note that this is the spread quoted by the specialist). For the most infrequently traded stocks the spread increased. The quoted depth declined by 48\% while the limit order book spread (i.e. the spread between the highest buy order and the lowest sell order) increased by 9.1\% (note that this result is in disagreement with previous studies). The tick size reduction had also an effect on transaction cost. Transaction costs for small  orders decreased even if this benefit is reduced for infrequently traded and low-price stocks.  Transaction costs for large orders either did not change (for frequently traded stocks),  or increased (for infrequently traded stocks). The authors were also able to track the behavior of different market participants. They found that after tick size reduction floor members were less frequently providing additional depth at existing limit order book prices while they were more frequently improving best prices. Contribution to displayed depth from floor members decreased by 35\% on average. Finally, limit order traders increased the ratio of cancelled limit orders to total limit orders by 6.2\%.

Huang and Stoll \cite{Huang01} compared two different market structures, namely the NYSE, which is an auction market with a tick size rule, and the LSE, which in the investigated period was a dealer market with no minimum tick size. They found that dealer market spreads are higher than auction market spreads, because in auction markets limit orders narrow spreads. Similarly depth is lower in auction markets because limit orders narrow the spreads, and these
spread narrowing  limit orders are small. Finally, in both markets they found evidence of clustering, which is the tendency for prices to fall on a subset of available prices. Quote clustering is highly correlated with spreads, while trade clustering is smaller in a auction market because limit orders break up quote clustering as they seek to gain priority.

Finally, Bacidore {\it et al.} \cite{Bacidore01} studied the tick size change from 1/16 to 1/100 at the NYSE (``decimalization'') by considering separately the NYSE and the Consolidated Quotation. They found that the NYSE's quoted spread fell by 30\% while non-NYSE spreads also generally decreased, albeit by relatively small amounts. Overall, the NBBO\footnote{The National Best Bid and Offer (NBBO) includes prices from all competing exchanges and refers to the price at the time of entry into the market.}
quoted spread fell by 30.7\%. Overall, NYSE quoted size fell by 70.5\% while non-NYSE average size declined by 26.9\% (in both cases the greatest percentage decrease occurred in the high-volume, low-price group). NBBO size fell by 61.5\%, which suggests that the non-NYSE markets became more likely to add to the NBBOÕs depth. The authors also investigated the overall shape of the limit order book. They found that displayed liquidity fell dramatically with decimal pricing, with a drop in displayed size of the order of 50\%. This effect appears to be greatest for low-priced stocks. As for the order properties,  decimalization had only a minimal effect on the relative fraction of market and limit  orders. However the average limit order size decreased by 33.4\% and
the average market order size decreased by 15.7\% after decimalization. Limit order traders became more aggressive after decimalization and the cancellation rate increased from 43\% to 53\%. The average time between quote updates declined and the total number of quotes across all markets increased;  this effect was stronger in non-NYSE markets. NYSE's share of quotes declined on average from 40\% to 34\%. The average number of NBBO  per stock per day doubled, while the number of quotes increased only by 27\% and the fraction of the average trading day that the NYSE is at both sides of the NBBO declined from 93\% to 82\%. All these results are consistent with a more competitive quoting environment.
The average transaction cost greatly increased after decimalization
\footnote{As in other studies, this is the average cost of trading a given
number of shares if the only liquidity in the market is the liquidity displayed in the book. 
Assuming that the midpoint of the spread is a proxy for the value of a security, 
the cost of displayed liquidity for an order is the product of the additional shares available 
in the limit order book at each price point times the distance of the price point from the spread
 midpoint summed through the number of shares in the order of interest. Dividing that sum 
 by the total number of shares in the order provides the per share cost of obtaining the displayed liquidity. This might be different from the actual cost of a large order obtained for example with order splitting.}.
 However, the effect is different depending on the volume. For all stocks, the cost of small orders decreased while the cost of large orders increased after decimalization. 
  
Taken together these studies indicate that a reduction in tick size (i) narrows the spread; (ii) decreases the quoted depth and the overall depth in the limit order book, i.e. the displayed liquidity decreases; (iii) modifies the order flow in a way that the rate of orders (and cancellations) per unit time increases, but their size becomes smaller; (iv) the transaction cost decreases for small orders, but increases for large orders (at least when the cost is computed by assuming that a large order climbs up the book); (v) creates a more competitive environment for liquidity provision inside a market (for example limit orders become more aggressive) and across different market segments. These two last aspects are probably important factors which contribute significantly  to the practice of order splitting and algorithmic trading. 

It is worth noting that the literature has given different sign to the relation between tick size and liquidity. On one side a small tick size reduces the spread, i.e. increases liquidity. On the other hand, it reduces depth and makes liquidity providers more reluctant to display their orders, with an effect of reducing liquidity. These contradictory results can be better reconcilied by considering that a complex concept such as liquidity is hardly captured by one metric.

\subsection{Tick size and price properties on longer time scales}

As said above, the literature on the relation between tick size and the statistical properties of prices is smaller than the huge literature on tick size and market microstructure. Here we review two recent papers from the Econophysics literature.

Onnela {\it et. al.} \cite{Onnela09} studied how tick size affects price return distribution.  They assumed a continuous price process that is discretized by the tick size. By using numerical simulations they found that the effect of discretization on return distribution due to tick size is negligible when the tick-to-price ratio is small, while it is significant when this ratio is large. 
Moreover, the proportion of zero returns appears to be much higher for stocks with a high tick-to-price ratio. They performed an empirical study by considering stocks traded at the same time in two exchanges with different tick sizes (NYSE and TSE). They observed that on average the proportion of zero returns increases as the tick size increases, but this is better accounted for by the tick-to-price ratio, whose variation explains roughly 69\% of variation in zero returns. Moreover 57\% of cases exhibit price clustering, such that the effective tick size deviates from the nominal tick size. 
In particular at the NYSE there is a strong preference for  even-eights. 

In a recent paper Munnix {\it et al.} \cite{Munnix10} considered two effects of tick size on price diffusion properties. The first is, as in the previous paper, on the return distribution and the conclusions are similar. The second is the effect of tick size on the estimation of cross correlation between price returns of two stocks. It is known \cite{Epps} that cross correlation between returns of two stocks declines when one reduces the length of the time interval used to compute returns. There has been several explanations for this, which is called the Epps effect, ranging from those based on learning to those on non-syncronous trading \cite{Reno,Toth}.  By using numerical simulations and analytical calculations authors of Ref. \cite{Munnix10} showed that even for synchronous time series, the discretization due to the tick size induces a distortion of the correlation coefficient toward smaller intervals. They then test their model on real financial data and they find that  the discretization effect is responsible for up to 40\% of the Epps effect. Moreover, the contribution of the discretization effect is particularly large for stocks that are traded at low prices. This highlights the importance of the tick-to-price ratio as compared to the absolute tick size. We will see below that more generally tick size affects the correlation properties of the second moment including its temporal autocorrelation.

\section{Tick size and price diffusion}

In this section we investigate the role of tick size in the statistical properties of price fluctuations, namely the return distribution, volatility clustering, and the subordination hypothesis.

\subsection{Return distribution}

The first question is how tick size change affects the distributional properties of returns. We have mentioned above two recent papers \cite{Onnela09,Munnix10} that discuss theoretically and empirically how the return distribution changes when tick size changes.

\begin{figure}[!htbp]
\begin{center}$
\begin{array}{cc}
\includegraphics[width=2.5in, height=2.5in]{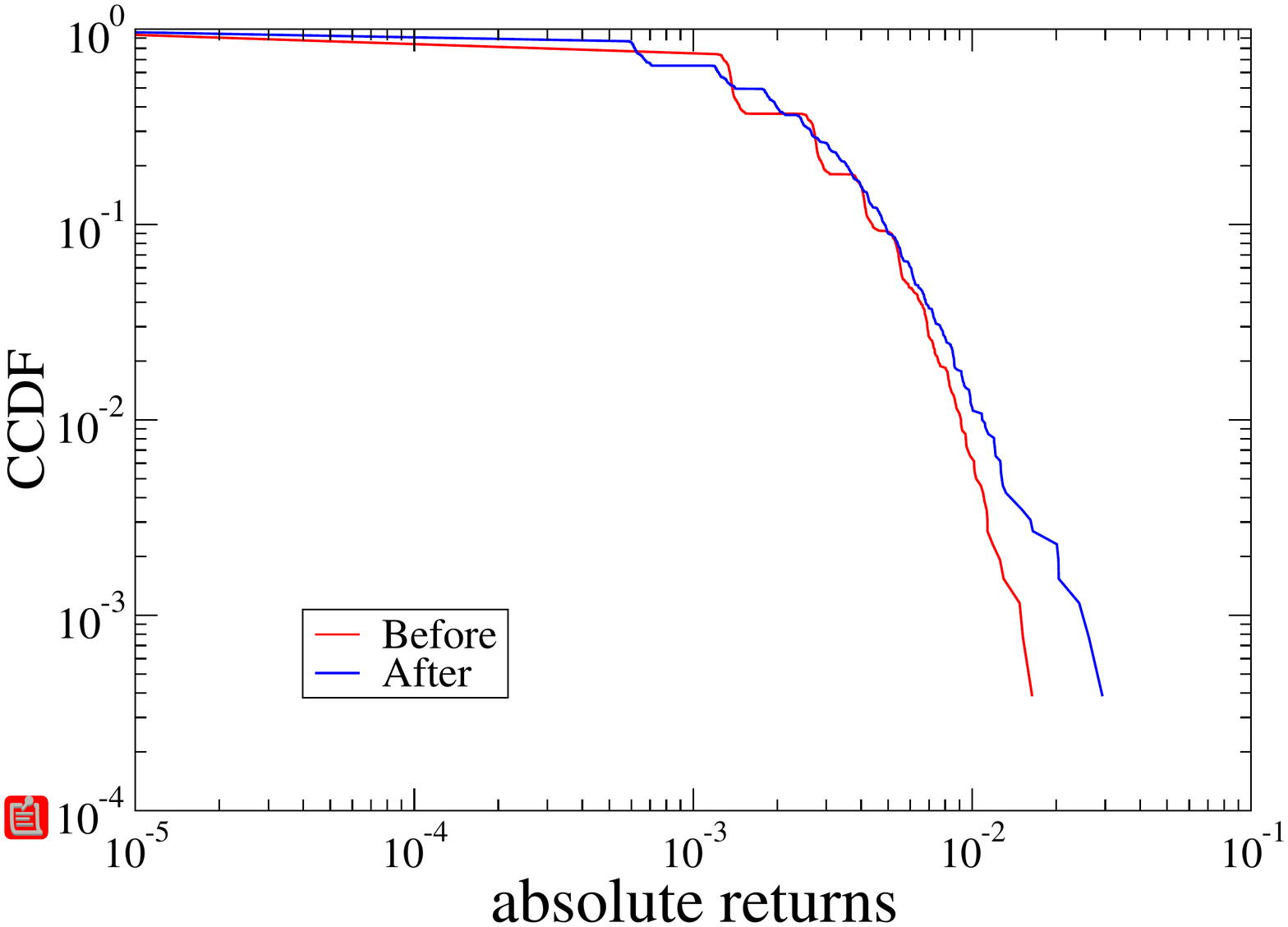} &
\includegraphics[width=2.5in, height=2.5in]{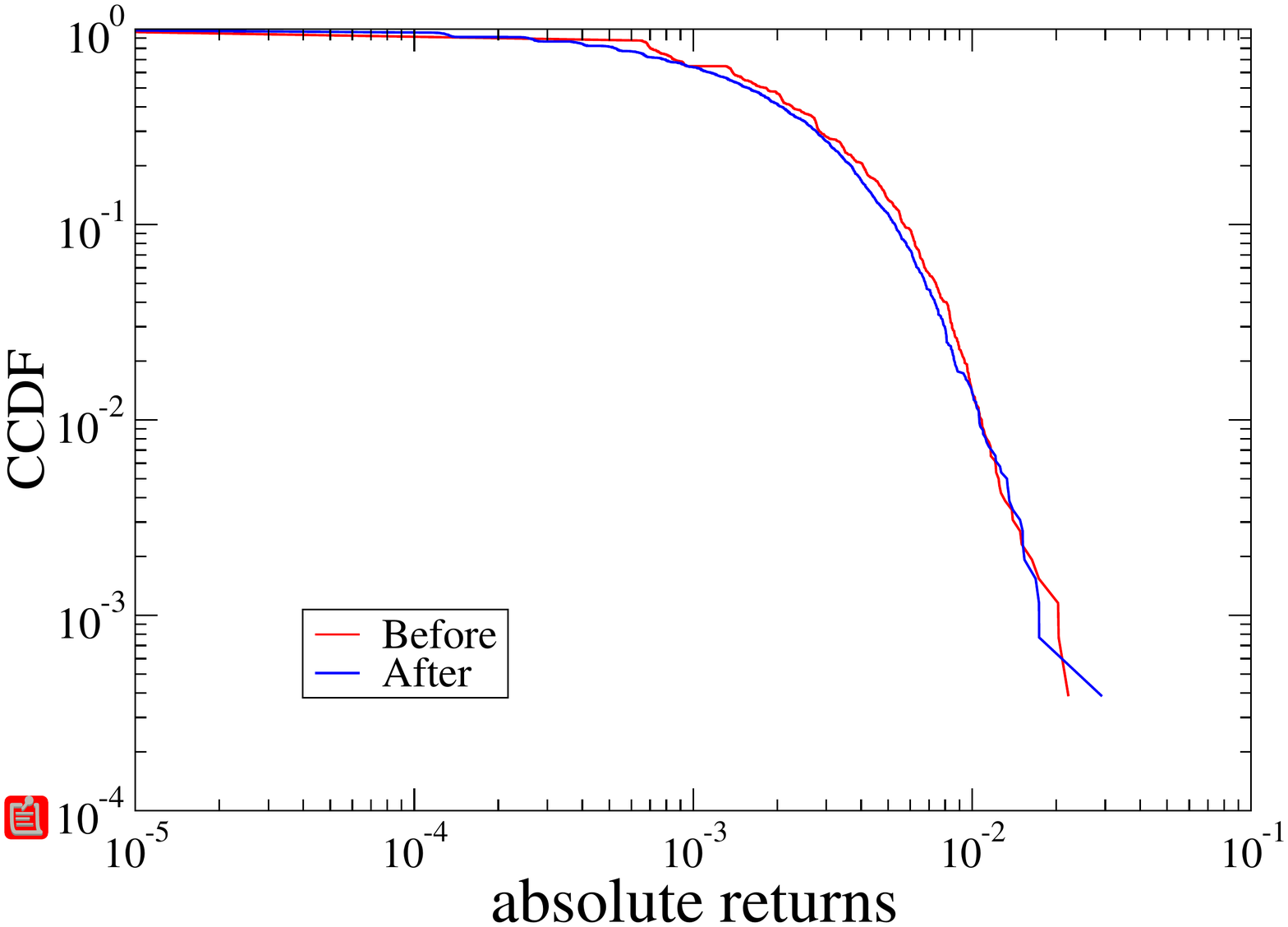}
\end{array}$
\caption{ 
 Complementary cumulative distribution function of the stock MRK 
 before and after the first tick size change (left panel), 
and the second tick size change (right panel). 
}
\label{EcdfMRKPlot}
\end{center}
\end{figure}

We use here a set of 5 high cap  stocks (KO, MRK, PEP, T, WMT) traded at the New York Stock Exchange (NYSE).  There have been two tick size changes in the NYSE. On June 24, 1997 the tick size changed from $1/8$ to $1/16$ of a dollar, and on January 29, 2001 the tick size changed from $1/16$ to $1/100$ of a dollar.  We consider short timescale log returns, namely, fifteen-minute returns. For each tick size change we consider two time intervals of length  $100$ trading days, one before and one after the tick size change. 
We consider absolute returns $|r|$ and we compute the complementary cumulative distribution function defined as $F_c(x)=\mathrm{P}(|r|\ge x)$. In Fig. \ref{EcdfMRKPlot} we show the complementary cumulative distribution function of 15 minute absolute returns for the stock MRK before and after the first tick size change (left panel), and the second tick size change (right panel). When the tick size changed from 1/8 to 1/16, both smaller non-zero returns and larger returns became more likely, i.e. 
 the distribution function of the absolute returns became more fat-tailed. On the other hand, when the tick size changed from 1/16 to 1/100, only smaller non-zero returns became more likely.
In this second case the distribution function of absolute returns did not become  more fat-tailed, but the number of small non-zero returns increased significantly. Similar results hold for all the other stocks.

\begin{figure}[!htbp]
\begin{center}$
\begin{array}{cc}
\includegraphics[width=2.3in, height=2.3in]{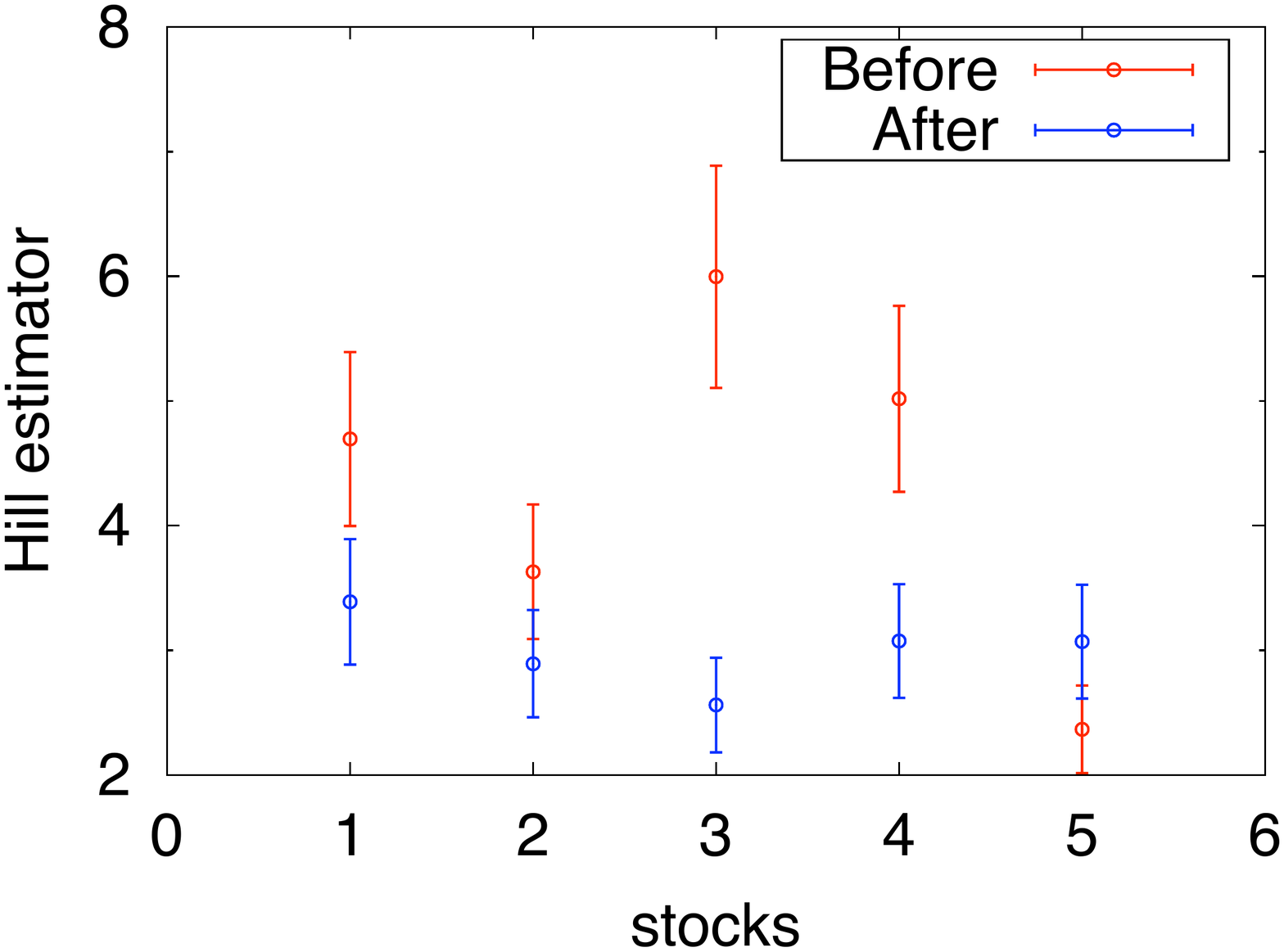} &
\includegraphics[width=2.3in, height=2.3in]{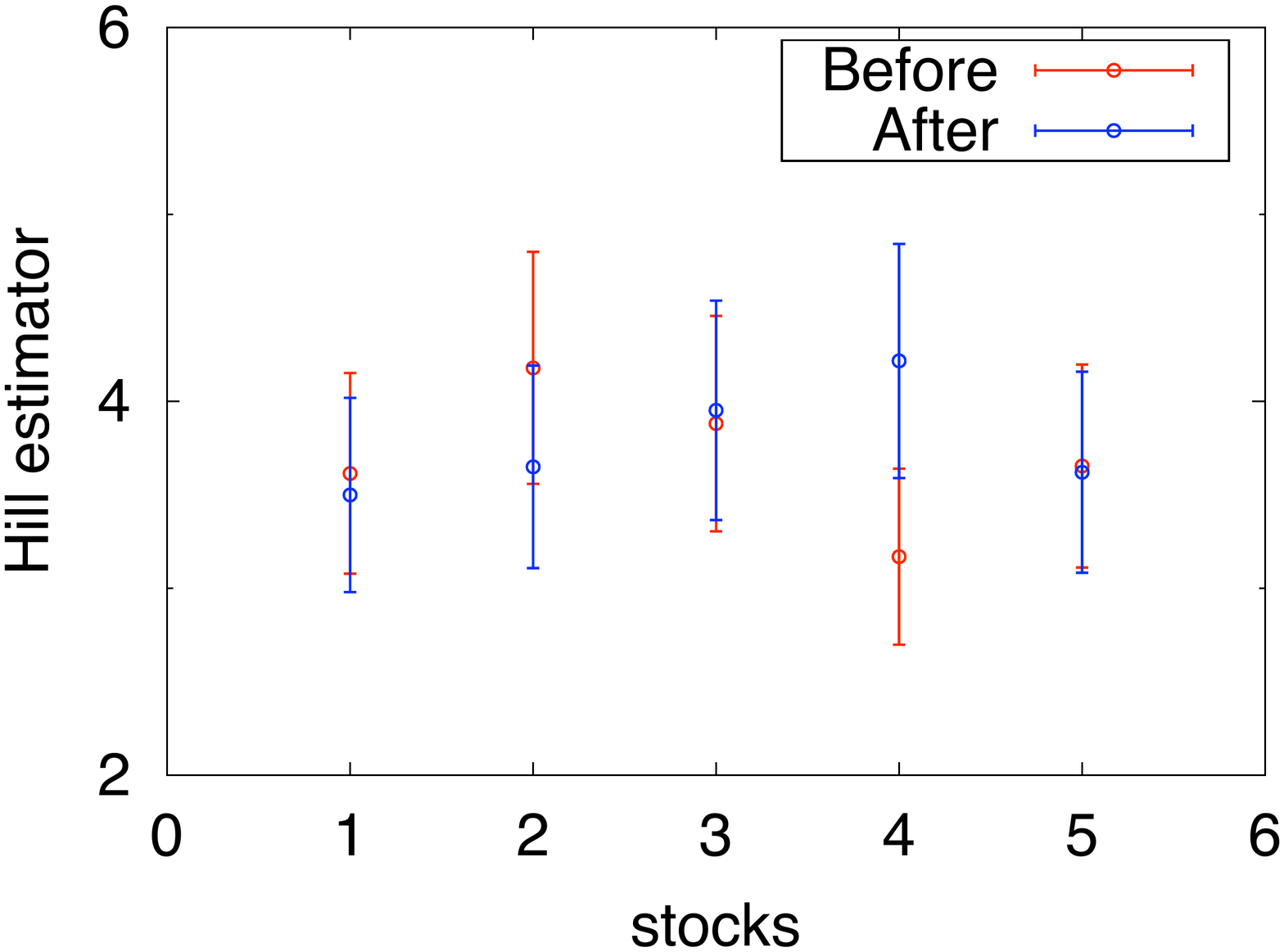}
\end{array}$
\caption{ Hill estimator for the tail exponent of the absolute return distribution before and after a tick size change from 1/8 to 1/16 (left panel) and from 1/16 to 1/100 (right panel). The stocks are alphabetically ordered and the error bars are 95\% confidence intervals.
}
\label{HillPlot}
\end{center}
\end{figure}

We use the probability of zero price change and the tail exponent to quantify the difference in the return distribution before and after a tick size change. We denote the frequency of zero returns by $p_0$. We define the difference in the frequency of zero returns just before ($p_0^-$) and just after ($p_0^+$) the tick size change, as $\Delta p_0=p_0^+-p_0^-$.  For each tick size change we perform a t-test of the null hypothesis that the mean of $\Delta p_0$ is statistically greater than or equal to zero. We obtain a {\it p} value equal to $0.0031$ and $0.0014$ for the first and the second tick size change, respectively. We can therefore reject the null that $\Delta p_0$ did not change or increase at a level of 1\% for both tick size changes. This means that in both cases the frequency of zero returns diminished. On average it reduced by 50\% on the first tick size change and by 70\%  on the second tick size change.  In other words, the price becomes less ``sticky'' when the tick size decreases. 
In order to study the tail properties of the return distribution, we make use of the Hill estimator $\alpha_H$ of the tail exponent. Figure \ref{HillPlot} shows the value of the Hill estimator of the 5 stocks before and after the two tick size changes. As before, we define the difference in the Hill estimator before and after a tick size change as $\Delta \alpha_H=\alpha_H^+-\alpha_H^-$ and  we perform a t-test of the null hypothesis that $\Delta \alpha_H$ is greater or equal than zero. For the first tick size change we can reject the null hypothesis at a level of 10\% ($p=0.06$), while for the second tick size change we cannot reject the null hypothesis ($p=0.62$). By using a shorter time period (50 trading days) on a larger set of 9 stocks we obtain a 5\% significant rejection of the null for the first tick size change, while again we cannot reject the null for the second tick size change. This confirms the intuition obtained from Fig. \ref{EcdfMRKPlot} that the tail of the return distribution changed after the first tick size change, but remained statistically the same in the second tick size change. The explanation of this difference remains an open issue.

In conclusion, our empirical results confirm that the return distribution can be different before and after a tick size change. In both cases the frequency of zero returns is higher for large tick sizes, while the behavior of the tails seems to be different in the two tick size changes.

\subsection{Volatility clustering}

The other important stylized fact of financial time series is volatility clustering. Roughly speaking, volatility clustering means that periods of high (low) volatility are more likely followed by periods of high (low) volatility. A method to quantify the degree of volatility clustering is through its autocorrelation function (ACF). It is commonly observed that the ACF of volatility decreases slowly to zero and is statistically different from zero for long time periods of the orders of months. Moreover, there is a consensus that volatility is a long memory process \cite{Ding93}, which means that for large lags the ACF $\rho(k)$ decays as a power law, $\rho(k)\sim k^{-\gamma}$ with $0<\gamma<1$. A long memory process lacks a typical time scale and can be characterized by the Hurst exponent $H$, which for long memory processes is given by $H=1-\gamma/2$. Here we investigate how the ACF and the (estimated) Hurst exponent changes when the tick size changes.  We consider the same stocks and time periods as in the previous section using the absolute price return for 15 minute intervals as a proxy of  volatility

\begin{figure}[!htbp]
\begin{center}$
\begin{array}{cc}
\includegraphics[scale=0.22]{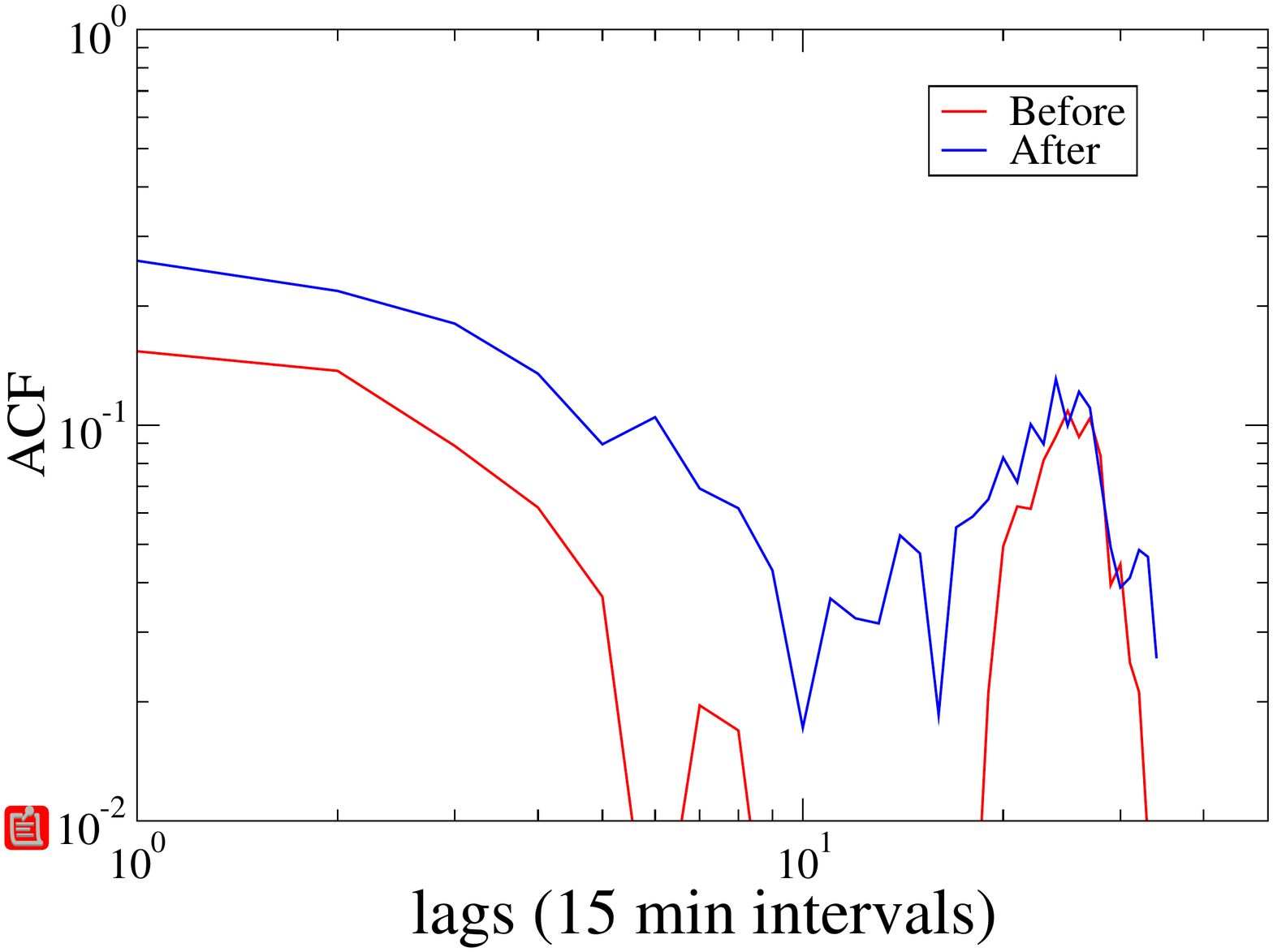} &
\includegraphics[scale=0.22]{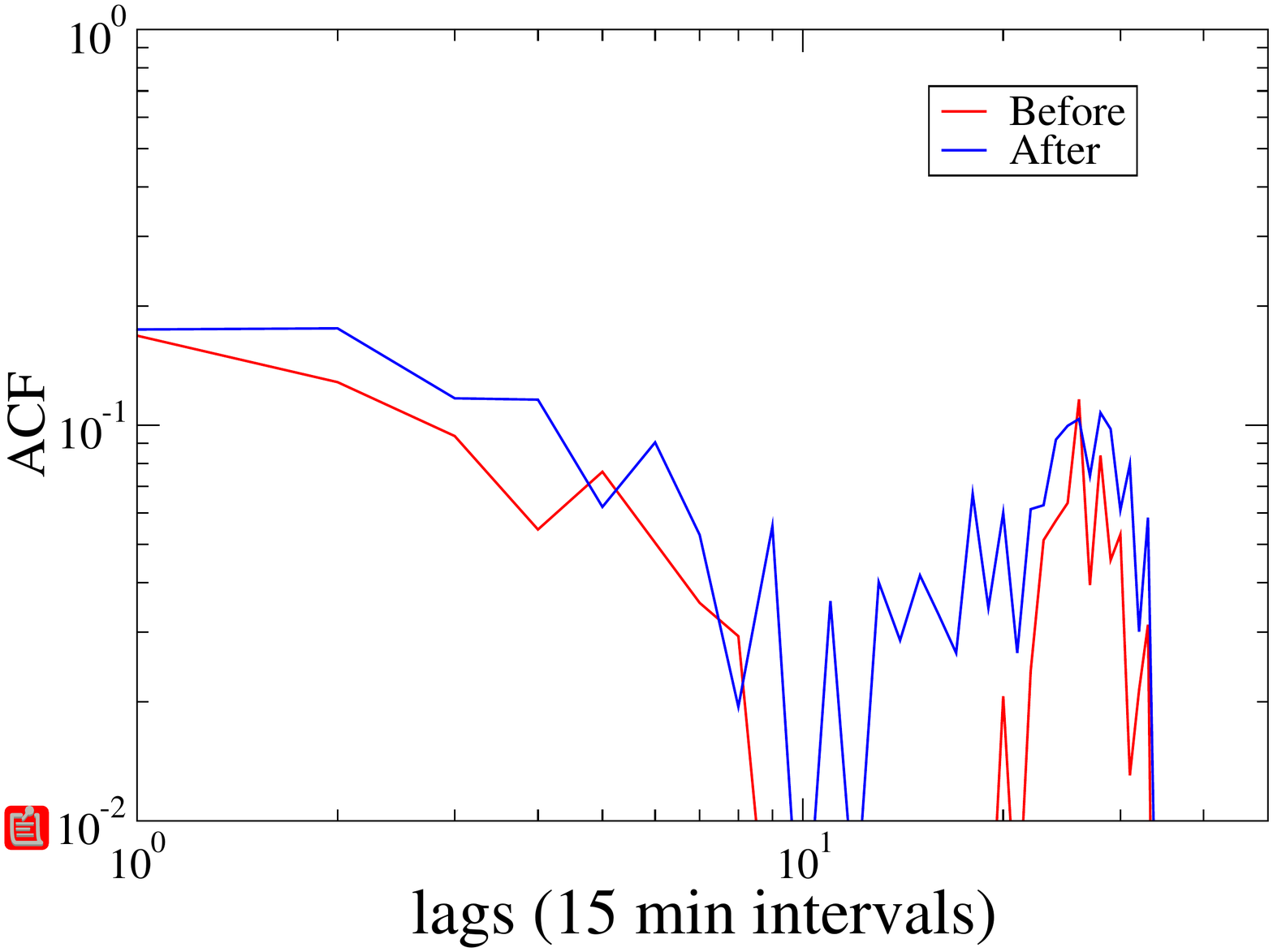}
\end{array}$
\caption{Autocorrelation function of 15 min absolute returns for the stock MRK in a 100 trading day period before and after the tick size change from 1/8 to 1/16 (left panel) and from 1/16 to 1/100 (right panel).}
\label{AcfMRKPlot}
\end{center}
\end{figure}

Let us denote the ACF of the absolute returns by $\rho(k)$, where $k$ is thenumber of 15 minute intervals. 
Fig.~\ref{AcfMRKPlot}Ê shows the ACF of volatility for the stock MRK for 100 days before and after the tick size change. In both cases the volatility ACF before the tick size change is smaller than the one after the tick size change. This suggests that volatility is less clustered (autocorrelated) when the tick size is large. In order to test  this conclusion more quantitatively  we define the difference in the autocorrelation function before and after the tick size change as $\Delta \rho(k)=\rho^+(k)-\rho^-(k)$, where $\rho^-(k)$ is the ACF
 just before the change, and $\rho^+(k)$ is the ACF just after the change. For each lag $k=1,..,4$ we perform a t-test of the null hypothesis that the mean of $\Delta \rho(k)$ is smaller than zero. Table \ref{AcfMRKPlot} shows the statistics and the {\it p} value. In all but one case we reject the null at the $1\%$ confidence level. In one  case  ($k=1$ in the second tick size change) the {\it p} value is $0.019$. This shows that the volatility becomes more correlated, i.e. more clustered, when the tick size is reduced.  In the next subsection we present a simple model explaining this effect. 
 
\begin{figure}[!htbp]
\begin{center}$
\begin{array}{cc}
\includegraphics[width=2.3in, height=2.3in]{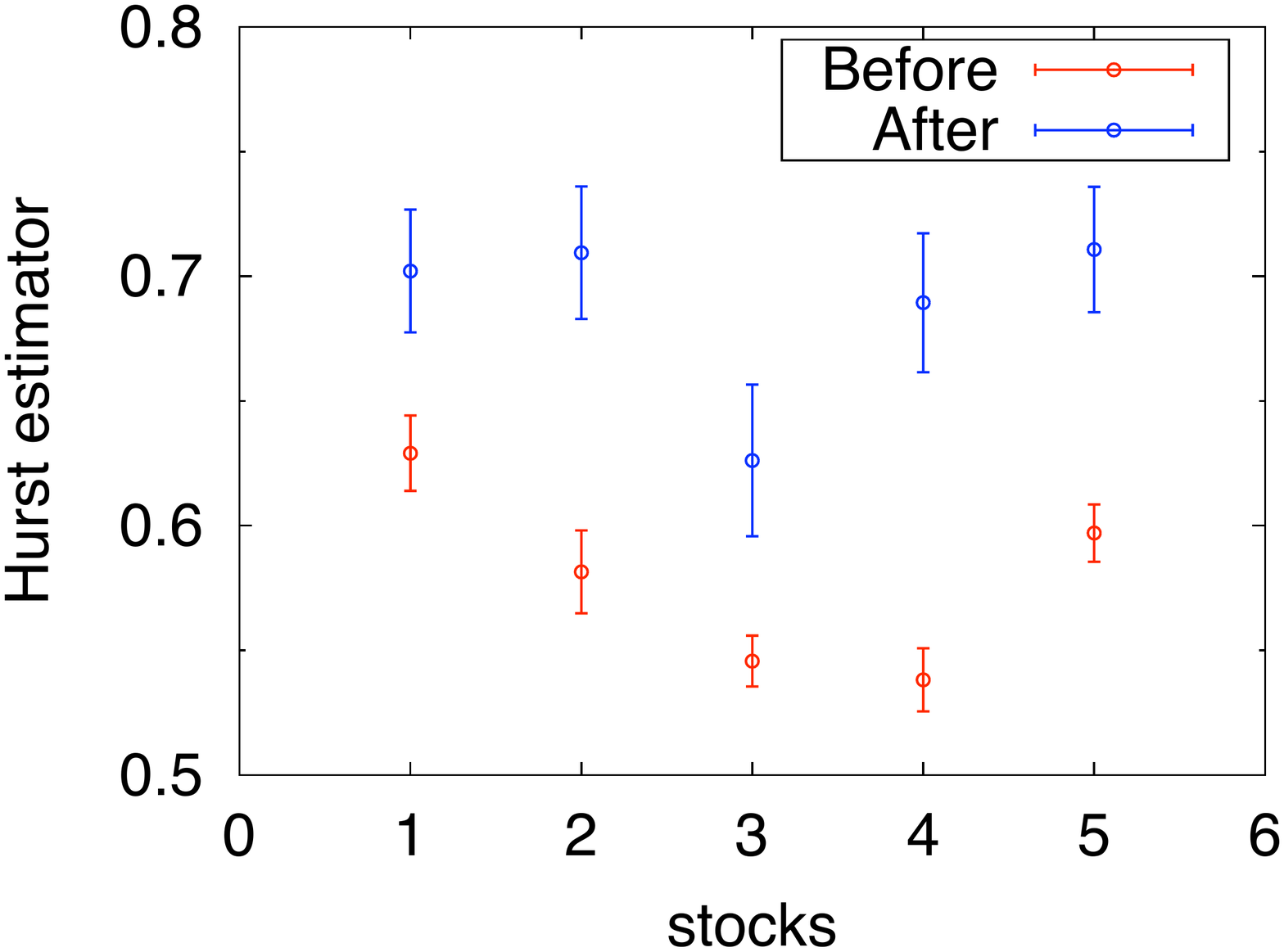} &
\includegraphics[width=2.3in, height=2.3in]{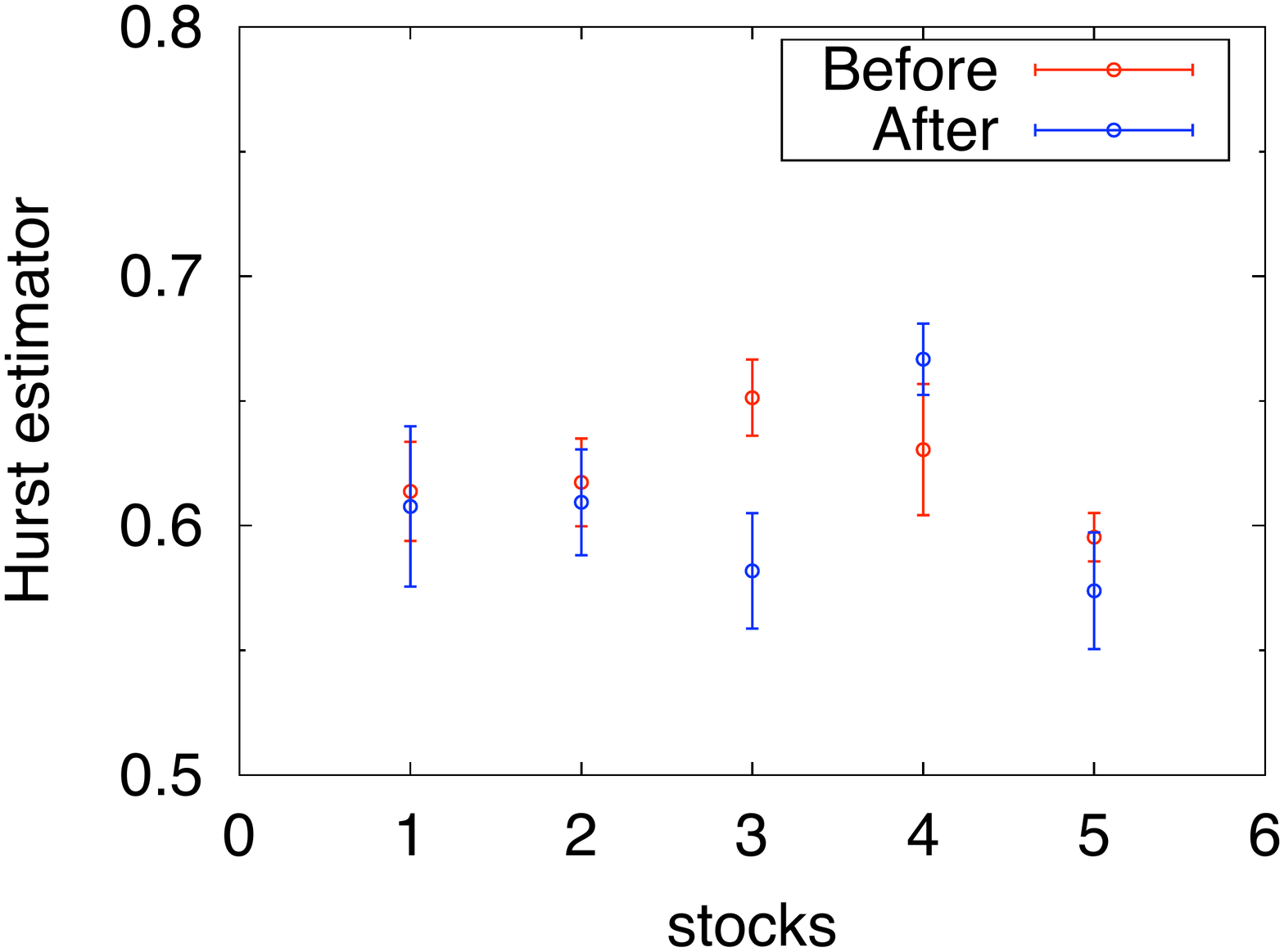}
\end{array}$
\caption{DFA estimator of the Hurst exponent of the absolute returns  before and after a tick size change from 1/8 to 1/16 (left panel) and from 1/16 to 1/100 (right panel). The stocks are alphabetically ordered and error bars are standard deviations.
}
\label{HurstPlot}
\end{center}
\end{figure}

 The previous analysis focused mainly on the autocorrelation for small lags. In order to investigate the change in the ACF for large lags before and after a tick size change we compute the Hurst exponent of absolute returns. It was suggested in \cite{Gillemot06} that the Hurst exponent of volatility becomes larger when the tick size becomes smaller. This is based on empirical evidence obtained by measuring the Hurst exponent in periods of about three years. Here we want to see if this effect persists on much shorter time intervals (5 months). The main problem is that we have shorter time series and therefore the estimation of the Hurst exponent is noisier. We estimate the Hurst exponent by using the detrended fluctuation analysis (DFA) \cite{Peng94}. Figure \ref{HurstPlot} shows the estimated  Hurst exponent of volatility for the two tick size changes.  The reduction of tick size from 1/8 to a 1/16  is associated with an increase of the estimated Hurst exponent, while in the second tick size change this phenomenon is not evident. A t-test of the null hypothesis that the variation $\Delta H= H^+-H^-$ in the estimated Hurst exponent is smaller than zero confirms this intuition .

 \begin{table}[ht!]
\begin{center}
\begin{tabular}{rrrrrrrrr}
  \cline{2-9}
 & \multicolumn{2}{c}{$k=1$} & \multicolumn{2}{c}{$k=2$} & \multicolumn{2}{c}{$k=3$} & \multicolumn{2}{c}{$k=4$} \\ 
  \cline{2-9}
 & t-stat & p-value & t-stat & p-value & t-stat & p-value & t-stat & p-value\\
 \hline 
tick size change \#1 & 3.7982 & 0.0096 & 5.8893 & 0.0021 & 5.6963 & 0.0023 & 7.0142 & 0.0011\\ 
tick size change \#2 & 3.0323 & 0.0193 & 7.4501 & 0.0009 & 3.9608 & 0.0083 & 8.7527 & 0.0005\\ 
   \hline
\end{tabular}
\caption{Result of a t-test of the null hypothesis that the difference  $\Delta \rho(k)$ of the autocorrelation function of 15 min absolute returns before and after a tick size change is smaller than zero.}
\label{AcfTTest_Long}
\end{center}
\end{table}

In conclusion, we showed evidence that a reduction in tick size leads to an increase of volatility clustering, measured as the autocorrelation function of 15 min absolute returns. In the next section we give some indication that this phenomenon can be explained with a mechanistic model of price dynamics. A more detailed explanation of this effect is presented elsewhere \cite{LaSpada10}. We found less evidence that the Hurst exponent changes when the tick size is modified. The increase of the Hurst exponent is clear after the first tick size reduction, while it is not observed during the second tick size change. This might be due to a poor estimation of the Hurst exponent. Mechanistic models \cite{LaSpada10} predict that the Hurst exponent should remain the same, but they also show that larger tick sizes have a lower {\it estimated} Hurst exponent. 

\subsection{A simple model}

We present here a simple mechanistic model that reproduces the increase in volatility clustering when the tick size is reduced. A more detailed explanation is given in \cite{LaSpada10}. Consider a price process described by a simple ARCH(1) process \cite{Engle82}
\begin{equation}
r_t=\sigma_tz_t~~~~~~~~~~~~~\sigma_t^2=\alpha_0+\alpha_1r_{t-1}^2
\end{equation}
where $r_t$ is the price returns on a given time scale and $z_t$ is a Gaussian noise term with zero mean and unit variance. We generated a return time series of length $2^{16}$ with parameters $\alpha_0=0.1$ and $\alpha_1=0.9$. We then construct the price time series $p$ by integrating the return time series\footnote{Note that we are considering additive rather than multiplicative returns. For the small time scales we are considering here the two give nearly equal results}. We assumed that this (unobservable) price is coarse grained by the tick size grid. Specifically, if the tick size is $\delta$ the observed price at the given time scale is $p_{obs}=[p/\delta] \delta$, where $[x]$ indicates the integer part of $x$. Finally, we construct observed returns from the observed price and we compute the autocorrelation function of absolute observed returns.

\begin{figure}[!htbp]
\begin{center}$
\begin{array}{cc}
\includegraphics[width=2.3in, height=2.3in]{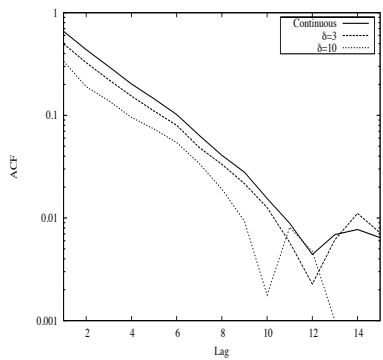} 
\end{array}$
\caption{Autocorrelation function of absolute returns of an ARCH(1) model and of a coarse grained version of it with different tick sizes $\delta$.
}
\label{ARCH}
\end{center}
\end{figure}

In figure \ref{ARCH} we show the autocorrelation function of absolute returns of the original process and of the discretized process. Note that the squared returns of an ARCH(1) price process are exponentially autocorrelated with a time scale $1/|\log \alpha_1|$. It is evident that coarse graining due to tick size reduces the ACF and that the larger the tick size, the smaller the autocorrelation function, similarly to what we observed in real data. This is a fully mechanistic model of the effect of tick size on volatility clustering. Other reasons can be at the origin of the empirically observed increase of volatility clustering after a tick size reduction. These may include microstructural effects, change in the strategic behavior of traders, etc. Our simple model and its extensions \cite{LaSpada10} show that part of  the effect could be due to purely mechanical reasons.

\subsection{Tick size and the subordination hypothesis}

The origin of fat tails and clustered volatility in price fluctuations is an important problem in financial economics   Although the cause is still debated, the view has become increasingly widespread that in an immediate sense both of these features of prices can be explained by fluctuations in volume, particularly as reflected by the number of transactions. The original idea dates back to a paper by Mandelbrot and Taylor \cite{Mandelbrot67} that was developed by Clark \cite{Clark73}. Mandelbrot and Taylor proposed that prices could be modeled as a subordinated random process $Y(t) = X(\tau(t))$, where $Y$ is the random process generating returns, $X$ is Brownian motion and $\tau(t)$ is a stochastic time clock whose increments are IID and uncorrelated with $X$. Clark hypothesized that the time clock $\tau(t)$ is the cumulative trading volume in time $t$, but more recent works indicated that  the number of transactions is more important than their size \cite{Ane00}.  Gillemot, Lillo, and Farmer \cite{Gillemot06} performed a series of shuffling experiments and showed that  neither number of transactions nor volume are the principal cause of heavy tails in price returns and clustered volatility. Specifically,  they compared returns (or volatilities) computed under different measures of time. They found that volatility is still very strong even if price movements are recorded at intervals containing an equal number of transactions (or volume), and that the volatility observed in this way is highly correlated with volatility measured in real time.  In contrast, when they shuffle the order of events, but control for the number of transactions so that it matches the number of transactions in real time, they observe a much smaller correlation to real time volatility.  

For the purpose of this paper we would like to stress the importance of the tick size in assessing the relative role of the subordination hypothesis in explaining fat tails and clustered volatility (in part discussed also in \cite{Gillemot06}). We consider here the 626 trading day period from Jan 1, 1995 to Jun 23, 1997, when the tick size at NYSE was 1/8 of a dollar, and the 734 trading day period from Jan 29, 2001 to December 31, 2003, when the tick size was a penny. We consider returns in real time, transaction time, and shuffled transaction time. Real time returns are simply 15 minute returns. Transaction time returns are obtained by considering time intervals containing an equal number of transactions. To make series with different time measures comparable we compute returns in transaction time by considering a number of transactions equal to the average number of transactions in 15 min. Finally, shuffled transaction time is obtained in the following way: We first measure the return of each transaction. We then shuffle the time series of individual transaction returns and we aggregate individual transaction returns so that we match the number of transactions in each real time interval.  In transaction time returns we are destroying any fluctuation of number of transactions while we preserve the temporal sequence of real trades. In contrast, in shuffled transaction time we preserve the fluctuations of trading activity (as measured by the number of transactions), but we destroy any temporal correlation between the amplitude of consecutive individual transaction price movements. If the subordination hypothesis is correct, then real time returns should be closer to shuffled transaction time returns than to transaction time returns.

\begin{figure}[!htbp]
\begin{center}$
\begin{array}{cc}
\includegraphics[scale=0.22]{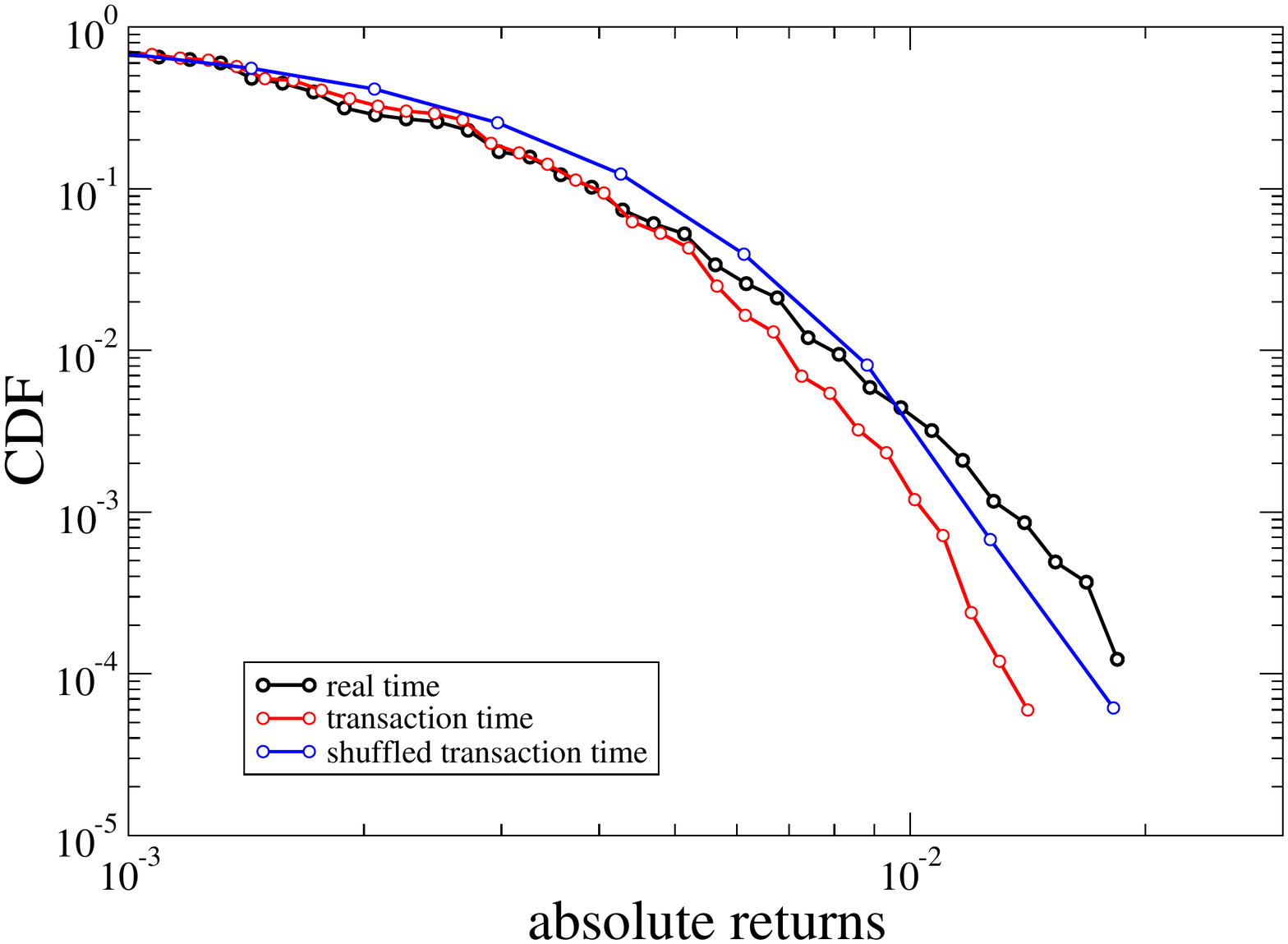} &
\includegraphics[scale=0.22]{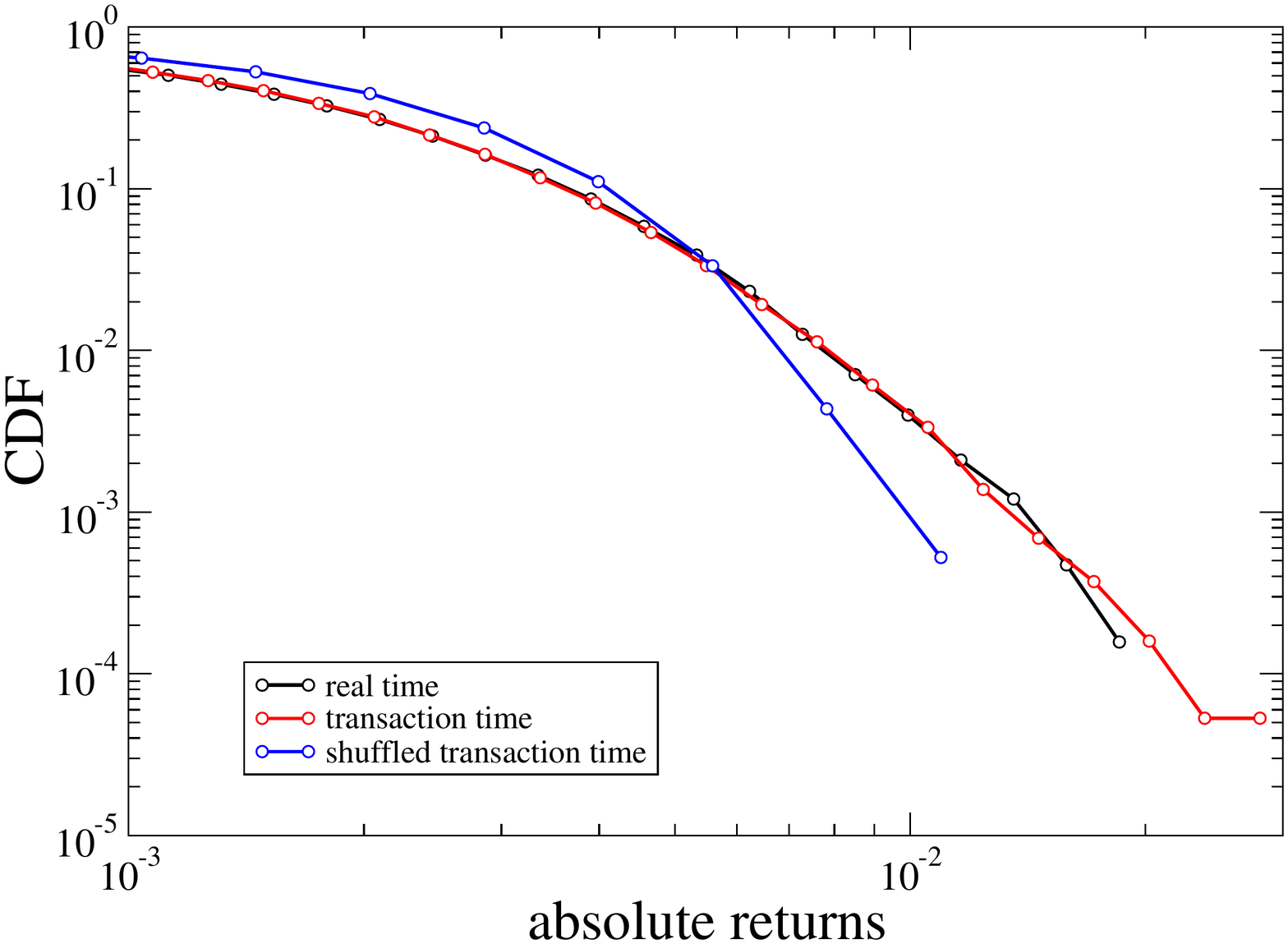}
\end{array}$
\caption{ 
 Complementary cumulative distribution function of absolute returns of the stock PG in the period when the tick size was 1/8 (left) and when it was a penny (right). We show the distribution of returns computed in real time (black), transaction time (red), and shuffled transaction time (blue).
}
\label{subordination}
\end{center}
\end{figure}

In figure \ref{subordination} we show the complementary cumulative distribution function of absolute returns of the stock PG in the two tick size regimes. When the  1/8 tick size was shuffled, transaction time returns are closer than transaction time returns to real time returns. This is qualitatively consistent with the subordination hypothesis. The period when the tick size was a penny shows a completely different pattern. The right panel of Fig. \ref{subordination} shows that in the small tick size regime the transaction time returns are closer than the shuffled transaction time returns to the real time returns. Clearly in this case the subordination hypothesis is not the main driver of fat tails. In other words, in order to reproduce fat tails of returns it is more important to preserve the temporal order of individual transaction returns than the fluctuations of their arrival rate. Since individual price returns are largely determined by liquidity fluctuations, these results indicate that when the tick size is small liquidity fluctuations are more important than volume (i.e. number of trades) fluctuations.  

\begin{figure}[!htbp]
\begin{center}$
\begin{array}{cc}
\includegraphics[scale=0.22]{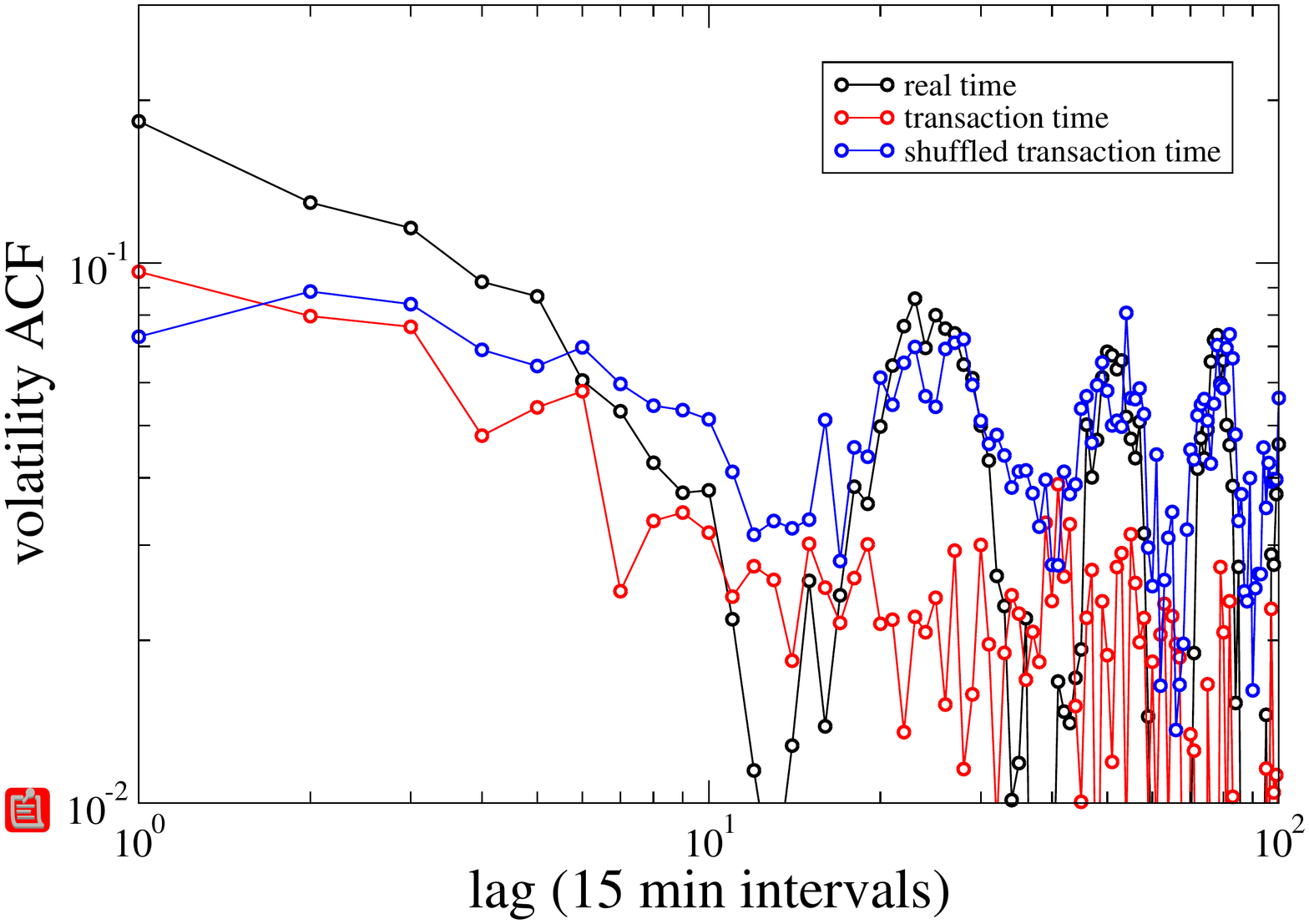} &
\includegraphics[scale=0.22]{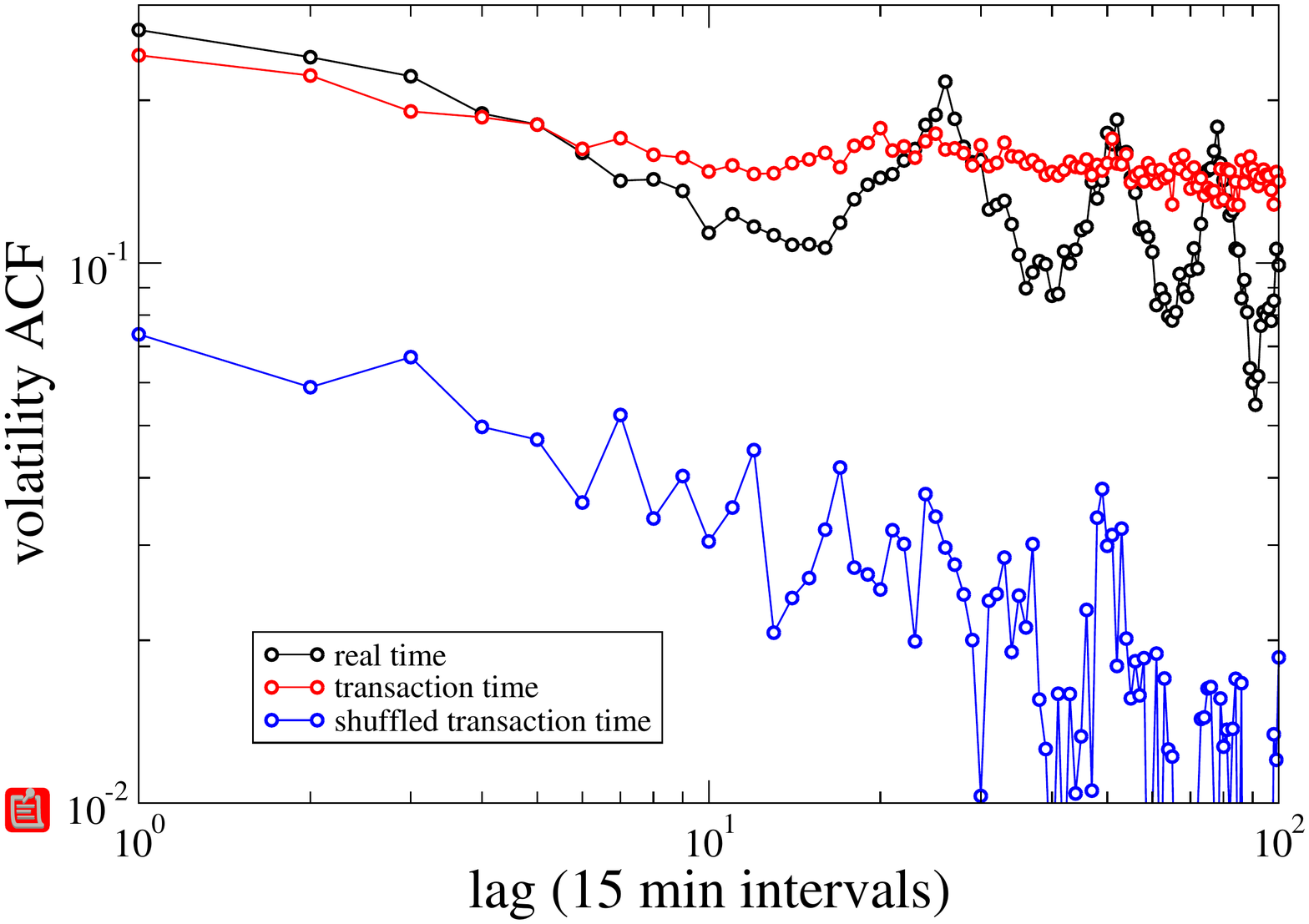}
\end{array}$
\caption{ 
 Autocorrelation of absolute returns for the stock PG in the period when the tick size was 1/8 (left) and when it was a penny (right). We show the autocorrelation of absolute returns computed in real time (black), transaction time (red), and shuffled transaction time (blue).
}
\label{subordinationvol}
\end{center}
\end{figure}

A similar conclusion can be drawn by investigating volatility clustering. Figure \ref{subordinationvol} shows the autocorrelation of absolute returns for PG under different tick size and time measures. 
When tick size is large (left panel), shuffled transaction time returns show volatility clustering that is close to the real one, while transaction time absolute returns are relatively less correlated. This again is in agreement with the subordination hypothesis. On the contrary, in the small tick size regime (right panel) the transaction time returns show volatility clustering very close to the real case\footnote{Note that real time and shuffled transaction time absolute return ACF show peaks due to the daily periodicity of trading activity. This is not observed for the transaction time absolute return ACF because of the way in which we have constructed the time series. Our discussion here refers to the global level of the ACF and not to the daily periodicities.}. Shuffled transaction time returns show a volatility clustering that is almost an order of magnitude smaller than the real case. Again, when the tick size is small, the subordination hypothesis plays a minor role in explaining volatility clustering.  We conjecture that the strong liquidity autocorrelation is an important driver of volatility clustering. In fact, most measures of liquidity, such as the spread \cite{Plerou} and distance between occupied levels in the order book (gaps) \cite{Lillo}, display strong autocorrelation, often consistent with long memory. Finally, note that the level of absolute return ACF in real time is much larger in the small tick size regime (right panel) than in the large tick size regime (left panel). This is in agreement with the analysis and the mechanistic model of the previous section.

In conclusion we have shown that the role of the subordination hypothesis in fat tails of returns and volatility clustering is strongly dependent on tick size. This is important also because many empirical studies of the subordination hypothesis have been performed on time periods of large tick size. Our study raises the question of the validity of these studies if they are applied to recent periods of small tick size.

\section{Conclusions}

In conclusion we have shown that tick size has multiple roles in influencing the statistical properties of price diffusion. As expected, and shown in the literature, tick size affects return distribution. When tick size is large (compared to the price) the price is more ``sticky'', i.e. there is an higher fraction of zero returns. It is less clear whether the tails of return distributions are affected by tick size. We have shown that tick size affects volatility clustering. A decrease in tick size leads to more clustered volatility. There might be many reasons for this effect that should be investigated. Here we have shown that a simple mechanistic model is able to  qualitatively capture this effect. Finally, tick size influences the relative role of volume (i.e. number of transactions) vs. liquidity fluctuations in explaining return distribution and volatility clustering. This relates to the importance of the subordination hypothesis in price diffusion. The original results presented here are preliminary in the sense that one should consider different time windows, use control windows to check for global trends, and consider different more tick size changes. 

The approach we have followed in this research is to directly study  the effect of tick size on price diffusion properties. We know that this effect is mediated by the price formation process, i.e. by the market microstructure. The long term goal of this research  is to understand how tick size affects market microstructure and how in turn microstructure affects price diffusion.

\printindex
\end{document}